# A Closer Look at the Light Induced Changes in the Visco-elastic Properties of Azobenze-Containing Polymers by Statistical Nanoindentation

Luca Sorelli,[1] Filippo Fabbri,[2] Jessy Frech-Baronet,[1] Mario Fafard,[1] Anh-Duc Vu,[3] Thierry Gacoin,[3] Khalid Lahil,[3] Yves Lassailly,[3] Lucio Martinelli,[3] and Jacques Peretti[3]

[1] Département de Génie Civil, Université Laval, G1V 0A6 Québec, QC (Canada)

[2] Institut d'Electronique Fondamentale, Université Paris-Sud/CNRS, 91405 Orsay (France)

[3] Laboratoire de Physique de la Matière Condensée, Ecole Polytechnique/CNRS, 91128 Palaiseau (France)

## ABSTRACT

The mechanical properties of azobenzene-containing polymer films are statistically measured by instrumented nanoindentation experiment in the dark and under illumination in the absorption band of the azobenzene molecules, with special emphasis on the creep behavior and recoverability. We use Dispersed Red 1 azobenzene derivatives, which remain in the stable trans isomer state in the dark and form a dynamical photo-stationary state between cis and trans isomer under illumination. Light induces a higher change in the film hardness than in the elastic stiffness, revealing the occurrence of a visco-plastic behavior of the film under illumination. Creep experiments performed at a constant load show a striking dissipative effect linked to the mass flowing under polarized illumination .

## 1. Introduction

In recent years, materials incorporating azobenzene derivatives have driven raising attention because of their singular mechanical response to a light stimulus.[1] The photo-mechanical properties of azo-materials stem from the isomerization of the azobenzene molecule between its *trans* (stable) and *cis* (meta-stable) configurations, subsequent to the absorption of a photon.[2] Various photomechanical effects have been demonstrated, leading to optical surface nano-patterning, as well as to optical actuation and optical control of motion and flows over a wide scale range, from macroscopic object sizes down to the molecular level.[3,4] These phenomena could potentially lead to new technological developments in various domains such as opto-electronics, ICT, data storage, micro-robotics, biology and healthcare, smart glasses, intelligent coatings, integrated optics, ophthalmic optics, etc.

In this perspective, the design and engineering of such materials require an in-depth investigation of the processes at the origin of the photo-deformation. However, although the various photomechanical effects in azo-materials have been widely studied since the first observation of a photo-induced deformation process in 1995,[5,6] the definitive understanding of the underlying microscopic mechanisms is still lacking. Addressing this challenging issue has stimulated many efforts both experimental[7,8,9,10,11,12,13] and theoretical,[14,15,16,17,18,19,20,21,22,23,24,25] but still the precise investigation of the photo-induced changes in the mechanical properties of the azo-materials has not yet been performed, especially in thin films. However a few studies are reported in the literature, all based on various experimental approaches and focussed on polymer materials incorporating different azobenzene derivatives.

Srikhirin and co-workers employed a quartz resonator technique to measure the elastic compliance of thin polymer films containing an azobenzene derivative with a long cis-isomer lifetime.[26] They show that the irradiation with low power density UV light, which depletes the trans isomer population, results in a slight increase (1.5%) in the polymer film stiffness, while the hardness tends to increase. Later on, Mechau et al. show that the elastic modulus of azobenzene containing polymer with side-chains decreases under illumination.[27] Indeed, UV irradiation can cause opposite effect on the elastic modulus for azobenzene-containing polymer depending on the possible presence of cross-links in their chemical structure.[28] By means of electromechanical spectroscopy Mechau et al. found that illumination with visible light, which may induce back and forth photoisomerization, causes a slight softening of the film, whereas ultraviolet irradiation results in an initial plasticization of the material, followed by its hardening.[27]

A widely used approach for probing the mechanical properties of thin films supported by a substrate is based on depth-sensing indentation measurement techniques.[29] In a landmark work, Karageorgiev *et al*. employed Atomic Force Microscopy to characterize a 380 nm-thick film of an azobenzene-containing polymer at very low penetration depth (10-20 nm).[30] Upon illumination, the elastic modulus is found to decrease by 74%, from 3.4 GPa to 0.9 GPa, as a consequence of the anisotropic fluidization along the direction parallel to the light polarization. The analysis of the creep response under constant load by fitting the Ting[31] formulation of the Maxwell viscosity yields an estimated light induced viscosity of a 4.1 GPa·s, which is three orders of magnitude less than the typical viscosity in the glassy state ($10^3$ GPa·s).[32] From a mechanics perspective, the choice of the Maxwell model, which cannot describe the creep behavior of a viscoelastic polymer may be questionable. Several other studies employed instrumented nanoindentation to characterize the mechanical changes in azobenzene containing polymers upon light excitation.[33] Nowicki *et al.* find that UV irradiation causes an increase in the indentation hardness and stiffness, while the viscosity decreases.[34] Notably, the changes in the mechanical properties induced by light are completely reversible. Moniruzzaman *et al* performed nanoindentation in copolymer films, containing azobenzene in the side-chain, having a thickness varying from 30 to 115 μm.[35] Upon UV irradiation, the elastic modulus is found to increase from 10% to 28% depending on the azobenzene concentration and this effect is mostly reversible after one day in darkness. Richter *et al* studied the effect of the illumination time and of the indent penetration depth (from 10 to 300 nm), under visible light illumination.[36] They find an increase in the material stiffness (about 140%) and hardness (between 100% and 300% depending on the penetration depth). Interestingly, in their experiments the modifications in the mechanical properties occur gradually during the first 120 s of irradiation, after which no change is observed. Surprisingly, the mechanical property changes are mostly irreversible 30 minutes after switching off the light. A recent nanoindentation study of Vapaavouri *et al* shows that the stiffness of the azobenzene-based polymer slightly reduces upon illumination, but more important the residual depth increases up to 80% depending on the applied strain rate.[37]

Table 1 summarizes the aforementioned results, which are apparently discordant. The reasons for such discrepancies can be imputed to the difference in the sample preparation, thickness and structure, the cross-linked structure of azo-molecules, the nature of light illumination in term of polarization and intensity, the diversity of the testing techniques. In spite of those possible mismatch sources, so far no study has provided a statistical characterization and analysis of the results dispersion for azo-polymer material.

In the present work, we use a customized, highly accurate nanoindentation technique to perform a statistical characterization of the light induced changes in the mechanical properties of azobenzene-containing polymer thin film. We also achieve creep measurements to characterize the dynamical response of the films under light activation and deactivation and to determine the reversibility of the viscous deformation.

| Reference | Mechanical property changes | | | Reversibility |
|---|---|---|---|---|
| | **Viscosity** | **Stiffness** | **Hardness** | |
| [26] | reduced by many orders of magnitude | +1.5% | n.a. | Completely recoverable |
| [36] | n.a. | +50 to 150% | +100% to 300% depending on the penetration depth | Mostly irreversible after 30' |
| [27] | n.a. | Slightly reduced | n.a. | Completely reversible |
| [30] | 4.1 GPa·s under light irradiation | -74% | n.a. | |
| [35] | n.a. | +9 to 28% | n.a. | Mostly reversible |
| [37] | n.a. | Slightly reduced | n.a. | |

**Table 1.** Published experimental results related to light-induced changes in the mechanical properties of azobenzene-containing polymer films.

## 2. Sample preparation and characterization

The material studied here, thereafter referred as PMMA-DR1, is a PMMA organic polymer backbone with Disperse Red 1 (DR1) azobenzene derivative molecules grafted as side-chains. It is prepared from a solution of 25 mg of a commercial compound (Sigma-Aldrich 570435) in 100 ml of dichloromethane. The material absorption spectrum (Fig. 1a) exhibits a broad peak centered at 487 nm corresponding to the visible absorption band of the DR1 trans isomer, which is the stable chromophore conformation. Illumination in the visible absorption band generates a photo-stationary equilibrium between the two isomer populations: absorption of a photon induces a conversion to the cis conformation while back conversion to the trans isomer occurs either by absorption of the same photon and/or by thermal excitation.[2]

The samples consist of 250-nm thick PMMA-DR1 films obtained by spin-coating the solution at 4000 rpm on a clean glass substrate. The photochrome concentration in the film is of about 1 molecule/nm$^3$. The photomechanical activity of the sample is checked by holographic inscription of a surface relief grating with light of wavelength 473 nm, in the absorption band of the azobenzene derivatives (Fig. 1a). A polarization pattern produced by interfering two linearly polarized beams of polarization directions tilted by ± 45˚ with respect to the *s* polarization is projected onto the back of the sample through the glass substrate. The photo-induced change in the surface topography of the film is measured in real-time by coupled shear-force and scanning near field optical microscopy (Fig. 1b).[12,38] The shear force microscopy image (Fig. 1c) shows before illumination a surface roughness of about 0.5 nm RMS, which is well below (at least by one order of magnitude) the values that could affect the nanoindentation measurements with respect to the experimental setup and the procedure used in this work.[39] When light is turned on, a surface relief grating forms with a 900 nm period given by the optical pattern inter-fringe. The photoinduced deformation amplitude kinetics during the first three minutes of illumination is shown in Fig. 1d, with the 3D representation of the surface pattern evolution. It is comparable with previous results obtained on similar films of the same material.[40]

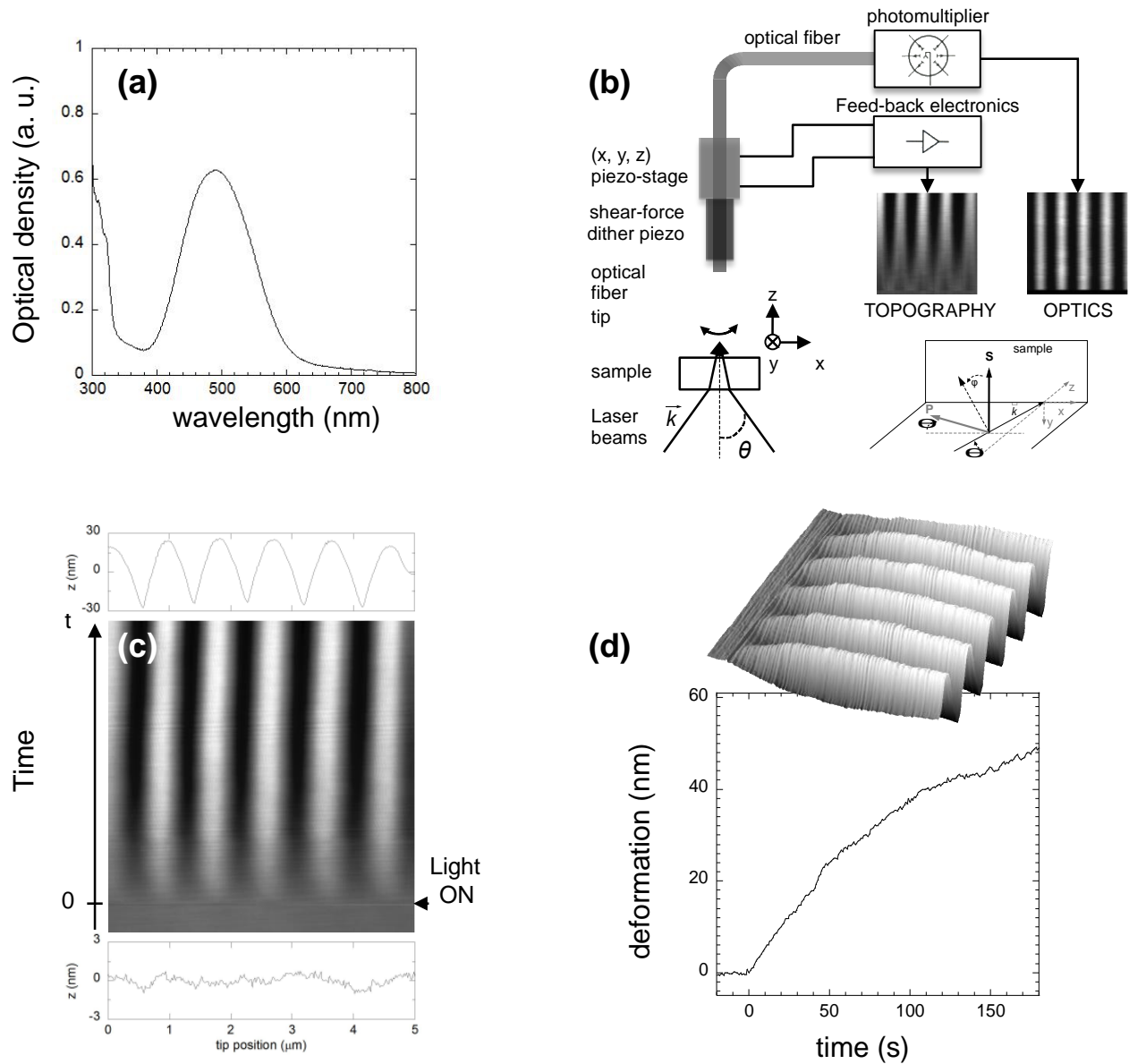

**Figure 1.** (a) PMMA-DR1 absorption spectrum, (b) Schematics of the experimental setup enabling simultaneous recording of topographical and optical images. The PMMA-DR1 film is illuminated by a polarized interference pattern. The angle φ defines the direction of the polarization field, (c) Shear-force image of the surface relief grating of 900 nm period inscribed during the projection of a polarized interference pattern on a 200 nm-thick azobenzene-containing PMMA thin film (wavelength λ = 473 nm, incident power density of 1 mW/mm²,

beam polarization orientation $\varphi = \pm45\degree$, incidence angles $\theta = \pm15.3\degree$). Surface profiles before illumination and after 180s of exposure show the surface roughness and the grating amplitude. (d) The variation of the surface relief amplitude as a function of time shows the photo-induced deformation kinetics.

## 3. Basics of nanoindentation

Nanoindentation has become a powerful technique to quantitatively measure the elasticity and hardness of soft polymer thin films.[41] During an indentation test, the force $P$ and the penetration depth $h$ of the indenter are simultaneously measured, while the indenter tip penetrates into the film (Figure 2a). A typical $P$-$h$ curve (Figure 2b) consists of a loading curve up to the maximum load $P_{max}$, followed by an unloading curve. Note that the loading curve profile is always convex in the case of conical indenters, if there is no friction on the tip-material interface. Among the several exiting types of indenter, we use the Berkovich-type indenter (Figure 2a), whose equivalent conical shape maximizes the tip sharpness and simple geometry is suitable for the interaction modeling. The Berkovich three-sided pyramidal diamond tip has an equivalent cone semi-angle $\theta$ of 70.3°. The analysis of the $P$-$h$ curve hinges on applying a continuum model[42,43] to derive the following indentation properties: the indentation modulus $M = S \sqrt{\pi} / (2\beta \sqrt{A_c})$ is directly related to geometrical parameters, the projected area of contact $A_c$ and the coefficient $\beta$ accounting for the slip on the indenter surface ($\beta$=1.034 for the Berkovich-type indenter [43,44,45]), the indentation hardness $H = P_{max} / A_C$, which is equivalent to the mean pressure supported by the sample under load. The projected contact area $A_C$ of the indenter-sample depends on the contact depth $h_c$ (cf. Figure 2a). The contact stiffness $S=(dP/dh)_{h=hmax}$ is the slope measured during the initial stages of the unloading curve. The Young modulus $E$ of an isotropic homogeneous material is estimated from the indentation modulus M as follows $M^{-1} = (1-v_i^2)/E_i + (1-v^2)/E$, where $E_i$ and $v_i$ are the elastic modulus and the Poisson's ratio of the diamond tip, equal to 1141 GPa and 0.07 respectively, while E and $v$ are the Young modulus and the Poisson's ratio of the material. Furthermore the area function $A_C$ $(h_c)$ of the diamond indenter is calibrated according to a standard procedure on a reference fused silica sample.[46] It is usually determined by the Oliver and Pharr method, which gives $A_c = 3\sqrt{3}h^2_c \tan^2\theta$, where $h_c = h_{max} - \varepsilon P_{max}/S$ with the geometric constant ε=0.72 for the conical indenter [46].

The viscoelasticity behavior is determined by performing a creep experiment, in which a constant load is applied and maintained during a given time, while the penetration depth increases. For instance, Figure 2c shows an experiment, where the load P is linearly applied over a time $\tau_L$, then is held constant over a time $\tau_H$, and is finally reduced to zero over a time $\tau_U$. The indentation creep coefficient C is generally determined by the ratio $h(\tau_L + \tau_H) / h(\tau_L)$, where $h(\tau_L + \tau_H)$ is the penetration depth at the end of the holding phase and $h(\tau_L)$ the penetration depth reached at the end of the loading phase. It is assumed that both elastic and plastic deformation processes occur during the loading of the indenter and elastic deformations are retrieved during the unloading curve.

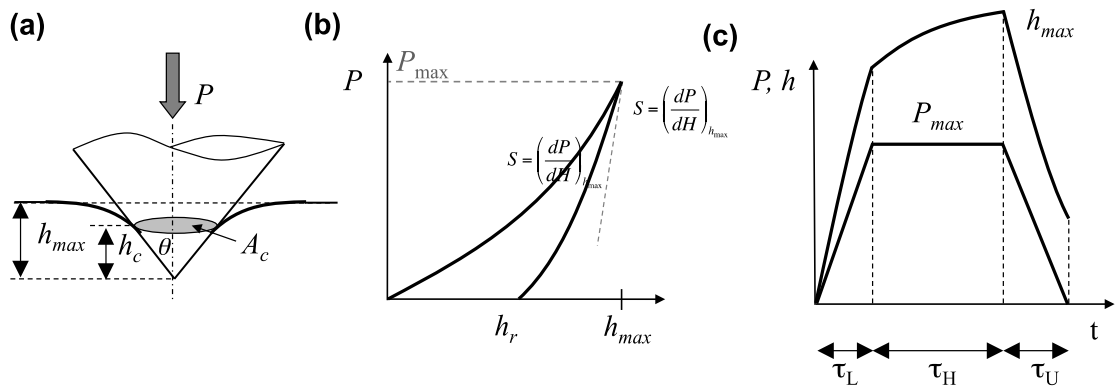

**Figure 2.** (a) Schematic view of an indentation test with a conical indenter showing various quantities used in the experiment: $h_c$ the contact depth, $h_{max}$ the displacement and $A_c$ and the projected contact area at maximum load, (b) Typical load P -penetration depth $h$, where $h_r$ is the residual deformation after indentation and S the contact stiffness measured at peak load. (c) Load profile P and displacement h of the tip versus time during a creep experiment.

The creep curves are interpreted by the classical Burgers viscoelastic model, described by 4 parameters ($E_1$, $E_2$, $\eta_1$, $\eta_2$), which corresponds to a combination of a Maxwell device with a spring and dashpot in series ($E_1$, $\eta_1$) and a Kelvin-Voigt device with a spring and dashpot in parallel ($E_2$, $\eta_2$), as shown in Figure 3-b. While the former represents the irreversible creep, the latter describes the reversible creep. In the case of a step function load application, the penetration depth is

$$\tau_L \leq t \leq \tau_H + \tau_L : \quad h^2(t) = \frac{\pi}{2} \frac{P_{max}(1-v^2)}{\tan(\theta)} \left( \frac{1}{E_1} + \frac{1}{E_2} \left( 1 - e^{-t/\tau_2} \right) + \frac{1}{\eta_1(1-v^2)} t \right) \tag{1}$$

Where $\tau_2 = \eta_2/E_2$ is the characteristic time of the reversible creep.[47]

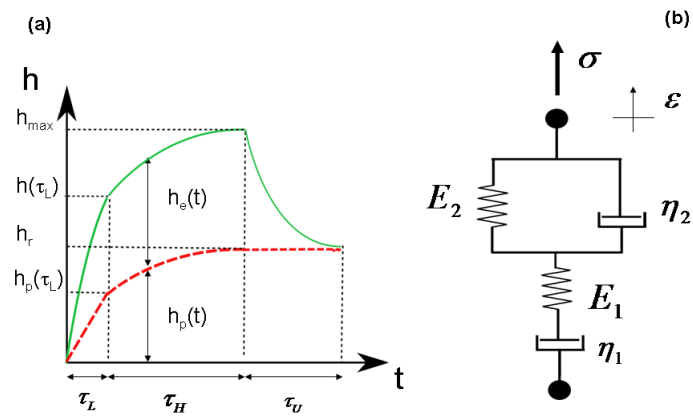

**Figure 3. (a)** Schematic penetration depth h-t curve showing the contribution of the plastic and elastic deformations: $h_r$ the residual deformation, $h_{max}$ the maximum deformation, $h_e$ the elastic

after loading and $h_p$ the plastic after loading. The red curve represents the irreversible plastic $h_p(t)$ variation. **(b).** The material viscous model considered in this study is the Burgers model which involves the viscosity coefficient $\eta_i$ and the elastic modulus $E_i$ of the material, where $\varepsilon$ and $\sigma$ are respectively the strain and applied stress.

The plastic deformation develops during the loading phase due to plasticity as well as during the holding phase due to the Maxwell irreversible creep (Figure 3-a). By assuming that the elastic modulus is constant during loading and unloading, the elastic penetration depth $h_e$ for a conical indenter can easily be deduced from: [48]

$$h_e = \sqrt{\frac{p}{2} \frac{P_{\max}(1-v^2)}{E_1 \tan(q)}} \tag{2}$$

One can then simply estimate the plastic deformation depth $h_p = h - h_e$ during the holding phase and unloading phase, with the hypothesis that unloading is purely elastic, as follows:

$$t_L + t_H > t > t_L: \quad h_p^2(t) = h_p^2(t_L) + \frac{p}{2} \frac{P_{\max}(1-v^2)}{E_1 \tan(q)} \frac{1}{h_1(1-v^2)}(t - t_L) \tag{3a}$$

$$t > t_L + t_H: \quad h_p^2(t) = h_p^2(t_L) + \frac{p}{2} \frac{P_{\max}(1-v^2)}{E_1 \tan(q)} \frac{1}{h_1(1-v^2)} t_H \tag{3b}$$

In this simplified model, we consider that the loading time $\tau_L$ and the unloading time $\tau_U$ are very short with respect the holding time $\tau_H$, i.e. $\tau_L$, $\tau_U \ll \tau_H$. As a consequence, since the loading and unloading processes are very rapid, we assume that the entire creep deformation occurs during the holding phase of constant load.

From an energy point of view, the energy dissipated during an indentation test can be estimated as follows:

$$W_p = W_{tot} - W_{el} - W_{el}^{creep} = \int_0^{h(\tau_L)} P_{loading}\,dh - \int_{h(\tau_L+\tau_H)}^{h(\tau_L+\tau_H+\tau_U)} P_{unloading}\,dh - \int_{h_p(\tau_L+\tau_H+\tau_U)}^{h(\tau_L+\tau_H+\tau_U)} P_{holding}\,dh$$

where $W_{tot}$ is the total energy applied from the external force; $W_{el}$ is the elastic energy recovered during the unloading; $W_{el}^{creep}$ is the recoverable elastic energy due to the reversible part of creep. The latter term can be estimated as the elastic part of the applied load energy during the holding phase. Finally, the percentage of total energy dissipated is simply estimated by $\gamma_p = W_p / W_{tot}$.

## 4. The experimental procedure

We utilize a special Anton-Paar UNHT nanoindenter, endowed with an active reference system in order to reduce the thermal drift to less than 1 nm over 500 seconds.[49] This feature makes this nanoindenter particularly suitable for accurately measuring small creep deformations over time. Experiments are conducted on an anti-vibration stabilized table, in a room with controlled temperature (20±1 ℃) and relative humidity (30±2.5%). Three experimental procedures are carried on. All of them are defined by different combinations of illumination and loading profiles as shown in **Figure 4**, where $\tau_L$ is the loading time, $\tau_H$ the holding time, $\tau_U$ the unloading time. $P_{max}$ and $P_2$ are the maximum and the final applied loads, respectively.

A series of preliminary tests are carried out to determine the optimal penetration depth for the PMMA-DR1 thin film. The effect of the substrate can be often ignored by following a common rule of thumb that the maximum indentation depth should be limited to less than 10% of the film thickness.[46,50] First preliminary tests performed with a penetration depth $h_{max} \approx 25$ nm, which corresponds to 10% of the film thickness requires a too low value of the load $P_{max}$ (about 5 μN) to give consistent results. The penetration depth $h_{max}$ is then increased step by step until the test repeatability is reliable, which is obtained for $P_{max} = 50$ μN and $h_{max} \approx 80$ nm. As a consequence, since the ratio between the penetration depth ($h_{max}$) and the film thickness is 40% in our study, our results must take into account the influence of the glass substrate. Let us outline that Ritcher *et al* found a threshold of 30% for thin polymer films.[36]

In Experiment 1 the **Load Profile 1** is applied to measure the visco-elastic properties of the sample in three illumination states: (i) before illumination, i.e. initial state (**Light Profile 1**); (ii) during illumination with visible light, hereafter termed as excited state (**Light Profile 2**); (iii) 30 min after illumination is turned off, i. e. relaxed state (**Light Profile 1)**. The loading is applied during the time $\tau_L = 8$ s. The load $P_{max} = 50$ μN is hold during the time $\tau_H = 120$ s. The unloading occurs during the time $\tau_{U1} = 8$ s until reaching zero load. The indentation measurements are repeated 25 times over a matrix of $5 \times 5$ indents with an inter-distance of 5 μm for each measurement.

Experiment 2 (**Load Profile 2)** allows characterizing the creep recovery of the sample before and during illumination (**Light Profile 1** and **Light profile 3**, respectively). The load profile and holding phase are identical to experiment 1. The unloading phase is performed in two steps: 1) a partial unloading taking a time $\tau_U = 5$ s until $P_2 = 2.5$ μN; this value of the load is maintained during a second holding phase with duration $\tau_H = 120$ s; 2) a final unloading phase to $P_2 = 0$, during a time $\tau_{U2} = 0.25$ s. Here again, the tests are repeated 25 times over a matrix of $5 \times 5$ indents with an inter-distance of 5 μm.

Experiment 3 aims at studying the transitory behavior of the azo-material when the light is turned on and off during the indentation (**Light Profile 4**). **Load Profile 1** is applied, with a longer holding time $\tau_H$ = 900 s. Starting from the beginning of the holding phase, the light is on after 60 s and off after 550 s.

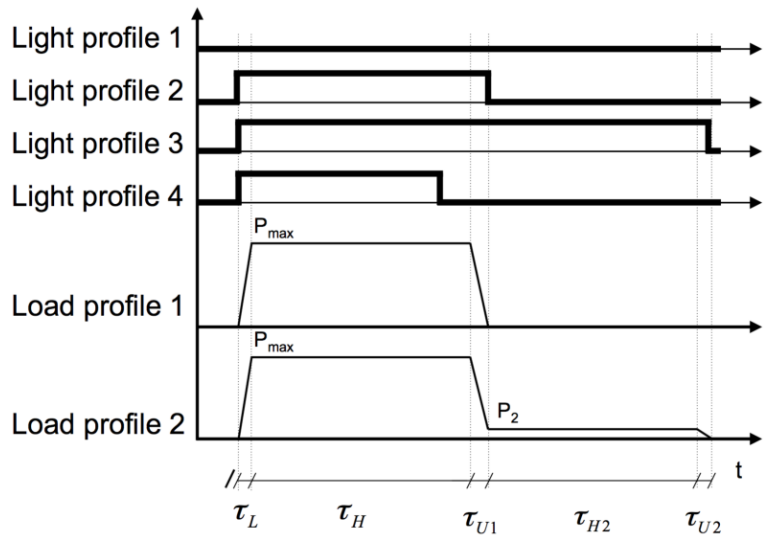

**Figure 4.** Load and concomitant illumination profiles applied in the various nanoindentation creep experiments.

## 5. Nanoindentation under illumination

As shown in **Figure 5**, the sample is illuminated through the glass substrate by means of a green laser beam, whose wavelength ($\lambda$ = 532 nm) lies within the DR1 molecule absorption band. The laser beam power density is about 2 mW/mm$^2$ and the angle of incidence $\theta$ is about 30° in air. The laser is linearly polarized; its polarization direction is tilted by about $\varphi \approx$ 45° with respect to s-polarization. The laser spot illuminates the sample through the glass substrate at the nanoindenter tip apex location.

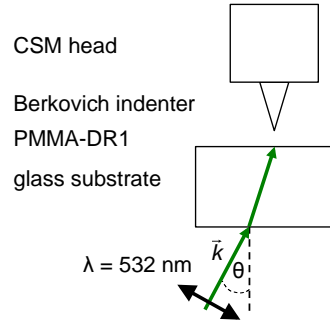

**Figure 5.** Experimental setup to perform nanoindentation under linearly polarized light.

First it is mandatory to check that the illumination has no effect on the glass substrate indentation. As shown in **Figure 6**, the indentation P-h curves are substantially unaffected by the light for a series of 50 tests. **Table 2** reports the mean values and the standard deviation of the indentation Modulus (M), Hardness (H) and Creep coefficient (C) of the glass substrate without and with optical excitation. Considering a Poisson's ratio of 0.3, the glass Young modulus is estimated to be 81 GPa, which is a typical value for glass.

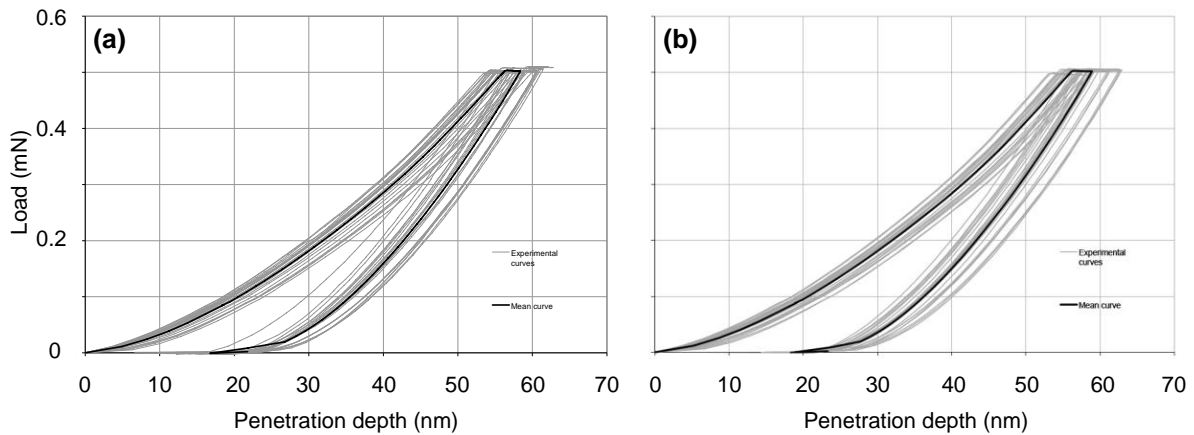

**Figure 6.** Load-penetration depth P-h curves for the glass substrate without (a) and with optical excitation (b). Black curves are averaged over 50 measurements.

| Indentation measurements | | Light OFF | Light ON |
|---|---|---|---|
| H [GPa] | Mean | 9.06 | 9.28 |
| | s.d. | 0.81 | 0.82 |
| M [GPa] | Mean | 88.99 | 89.53 |
| | s.d. | 4.33 | 4.19 |
| C [%] | Mean | 4.65 | 3.61 |
| | s.d. | 1.49 | 2.50 |
| $\gamma_p$ [%] | Mean | 0.39 | 0.38 |
| | s.d. | 0.04 | 0.06 |

**Table 2.** The indentation quantities estimated from the P-h curves for the raw glass substrate with and without optical illumination (s.d. = standard deviation)

Let us now start to perform statistical series of nanoindentation measurements following the procedure defined in the Experiment 1 step. The initial state prior to illumination is characterized by the load-penetration (P-h) and penetration depth versus time curves shown in **Figure 7**(a-b). The same measurements achieved during illumination and in the dark 30 minutes after illumination yield the curves c-d and the curves e-f, respectively. The mean P-h (a) and h-t (b) curves are compared for the 3 considered states in Figure 7(g-h). The following conclusions can be drawn from the comparison between the three states : i) Illumination clearly increases the maximum penetration depth $h_{max}$ by 30 %, ii) The relaxed state has the same quantitative mechanical response than the initial state, which asserts the reversible character of the mechanical changes induced by the photoisomerization of azobenzene molecules. iii) Unlike the initial state, the excited state reveals a large dispersion in the P-h and h-t curves. It cannot be attributed to the light intensity fluctuations since such dispersion is not measured when achieving nanoindentation on the raw glass substrate under illumination. More likely it is related to the anisotropic displacement of the polymer chains under the light polarization.

In the second experimental procedure (Experiment 2) the load is discharged from $P_{max}$ to $P_2$ during the time $\tau_{U1}$ and the load $P_2$ is maintained during the time $\tau_{U2}$ (cf. Figure 4b). The creep recovery curves (h-t) are shown on **Figure 8(a)** in the initial and excited states. The asymptotic behavior of the penetration depth, common to both states expresses the partial recovery of the viscous deformation. Indeed upon unloading, the penetration depth suddenly reduces due to the elastic deformation, followed by a viscous reduction of the penetration depth. The illuminated material shows a creep recovery smaller than the one measured in the initial state. Moreover, the characteristic time of the creep recovery is much shorter than that of the creep under loading.

Under illumination the material has developed a greater irreversible creep during the holding phase, which cannot be recovered after unloading. Therefore, since the permanent deformation, which is the irreversible penetration depth upon unloading increases under illumination, we can conclude that illumination favors the visco-plastic behavior of the material.

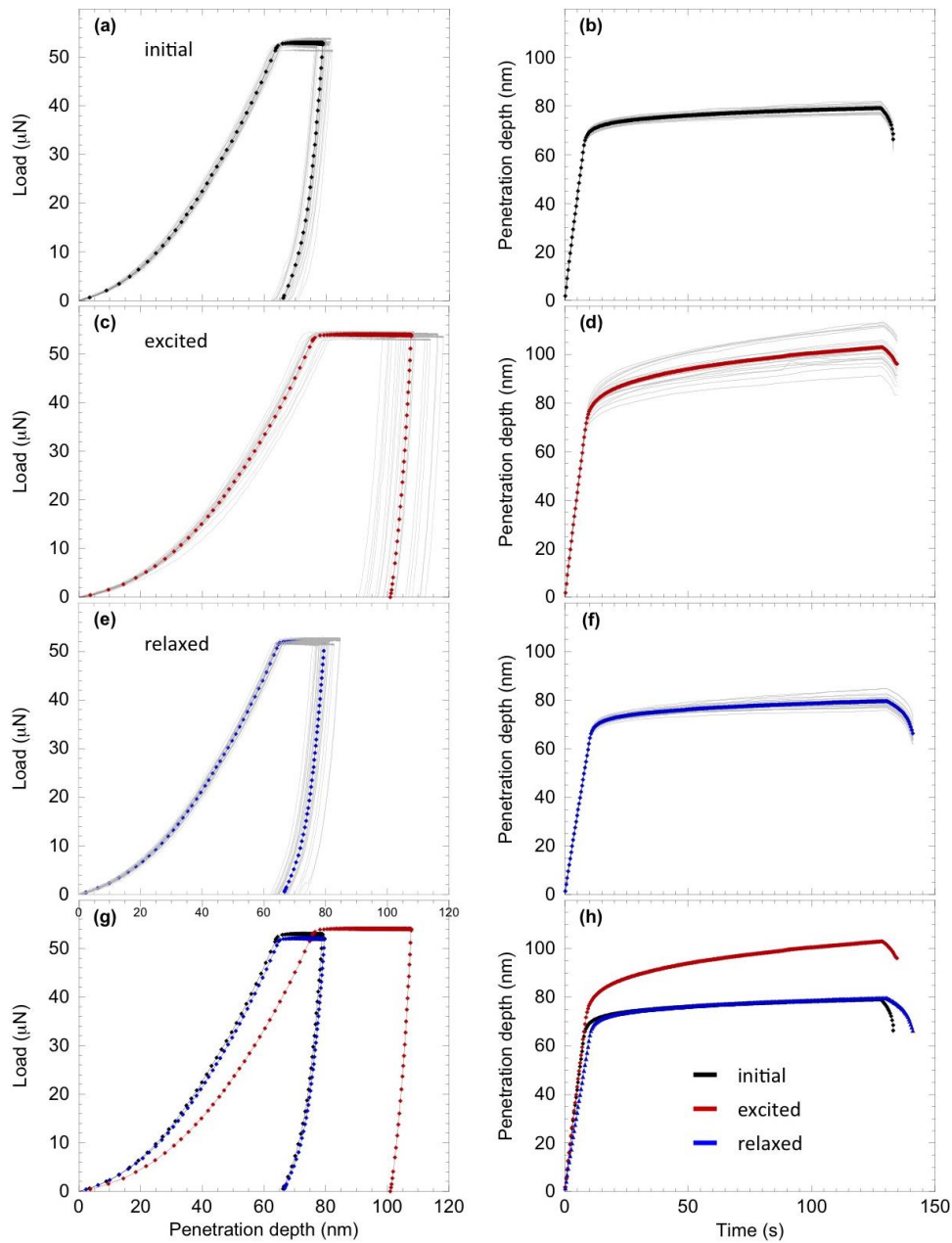

**Figure 7.** P-h curves and h-t curves measured on PMMA-DR1 with $P_{max} = 50\,\mu N$ without illumination (a,b), during illumination (c-d) and after illumination (24h in the dark) (e-f). The average curves are drawn in dark. (g-h) Average P-h curves (a) and average h-t curves (b) for the azobenzene polymer sample before, during and 30 minutes after illumination.

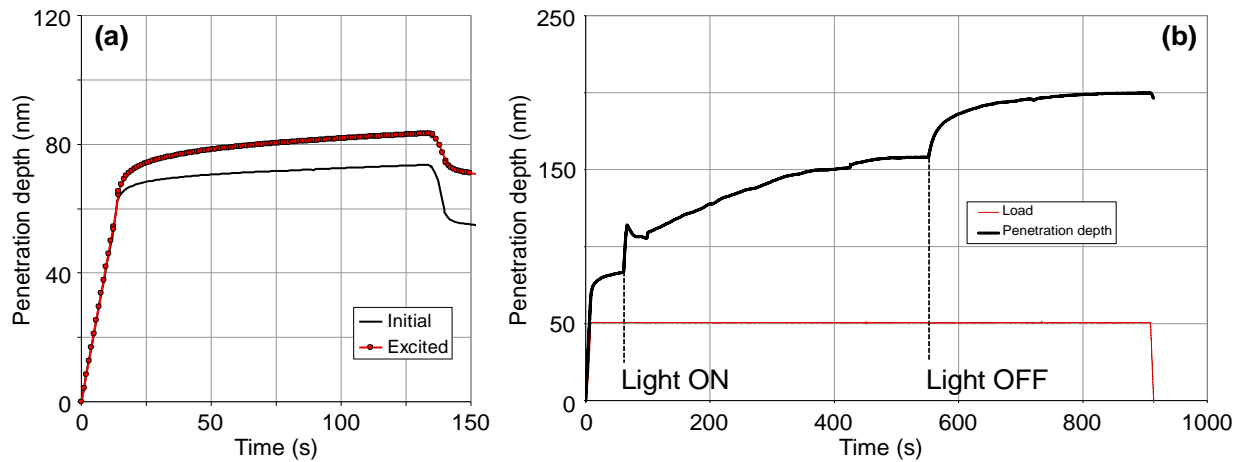

**Figure 8. (a)** Average creep recovery curves of the DR1 azobenzene polymer sample before (black curve) and during (red curve) illumination from $\tau=0$ up to $\tau_{U2}$ (b). **(b)** Switching on/off the light during the creep experiment (in red the load profile $P_{max} = 50.6$ μN).

The last experimental procedure (Experiment 3) aims at measuring the dynamics of the change in the material mechanical properties due to light excitation. For this purpose, we measure the penetration depth of the indenter when the light is successively switched on and off, while the the film is submitted to a constant load of 50.6 μN (**Figure 8b**). Turning on the light 60 seconds after holding the load constant, the penetration depth shows a sudden rise hinting to a very rapid softening of the material within less than 1 second. It is likely that this transitory feature is directly related to the building up of a photostationary equilibrium between trans and cis isomers, which occurs within the same timescale. Indeed, the photostationary equilibrium is reached concomitantly with the completion of the photoexpansion of the azo-polymer,[12,13] which was shown to occur for the same light exposure. The instantaneous increase in the penetration depth (~40 nm) over 4 seconds is approximately the same increase in the penetration depth over 100 seconds obtained in the measurements performed in *Experiments 1 and 2*. One can roughly estimate that turning on the illumination leads to a decrease in the polymer viscosity by 2 orders of magnitude and the occurrence of the fluid state. However the most striking effect is the rapid rise of the penetration depth when the light is turned off, while the creep rate stays similar to that before illumination. The sudden absence of light stimulus releases the uniaxial stress applied to the molecular chains imposed by the light polarization, which can now freely reorganize in an isotropic way. The chain orientation under light-induced stress creates a spatial order characterized by a decrease in entropy, which is broken when the light is turned off. The resulting increase in entropy might contribute to maintain a fluidization in the film.

## 4. Discussion

### 4.1 Light induced modification of the indentation parameters

**Figure 9** compares the mean values of the indentation parameters (modulus M, hardness H, and creep coefficient C) for the initial excited and relaxed states. The value of the indentation modulus M, which reflects the stiffness of the material, is quite low. M increases of about 8% when light is turned on and returns to the initial value when light is turned off. This result is in agreement with previous studies, which observe a slight increase in the material stiffness under illumination.[26,35,37] On the other hand, indentation hardness reduces of about 38% when light is turned on and gets back its initial value when light is turned off. The most drastic variation concerns the indentation creep coefficient C, which rises of 98% under light illumination. All the indentation quantities M, H, and C are completely reversible when light is turned off. The standard deviation of the coefficients remains quite low in the initial and relaxed states (4-5% for M and H, and 10% for C), but it significantly increases under illumination (7% for M and H, and 14% for C). The increase might be of topological origin such as the anisotropic disentanglement of polymer chains triggered by cyclic photoisomerisation of azobenzene molecules. Notably, the reduction in H and the increase in C suggest that the material behaves in a more visco-plastic fashion under optical excitation. In other words the optically illuminated PMMA-DR1 film develops more plastic deformation favoring a molecular flow driven by the reduced viscosity of the film. Thus the light absorption of DR1 molecules leads to generate an anisotropic stress well beyond the visco-elastic regime, as if the plastic barrier for the visco-plastic deformation is lowered by the molecular chain motions.

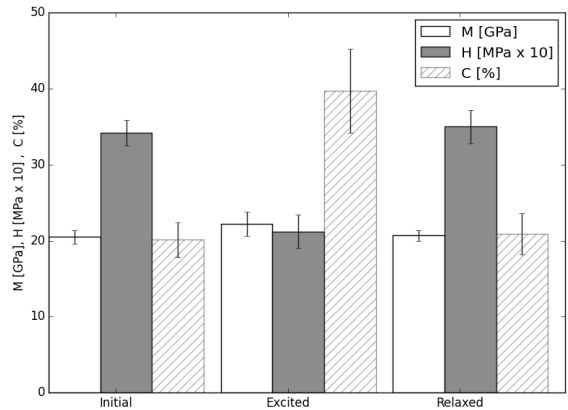

**FIGURE 9.** Histograms of the indentation Modulus, Hardness and Creep for the initial, excited and relaxed states (standard deviations are indicated by the error bars) for 250 nm thick azobenzene PMMA film.

Several models have been developed to extract the intrinsic elastic modulus of thin films from the composite film/substrate modulus value obtained from indentation tests on coated systems. To estimate the modulus $E_f$ of the azobenzene containing polymer film, we use a closed solution[51, 52, 53, 32] for high contrast between the stiffness of the film and the substrate. In essence, the composite shear modulus is:

$$G_c^{-1} = \left(1 - I_0\right)\frac{1}{G_s + F\, I_0\, G_f} + I_0 G_f^{-1}$$

which links the effective measured composite shear modulus $G_c$ to the shear moduli of the film $G_f$ and of the glass substrate $G_s$, noting that $G_c = E_c / 2(1+v_c)$ can be derived by the measured indentation modulus M directly as explained in section 3. F is a constant, which has been estimated by finite element computation to be $0.026$[51]. In this expression, the weight function $I_0$ reads:

$$I_0 = \frac{2}{\pi}\arctan\left(\frac{d}{a}\right) + \frac{1}{2\pi\left(1\text{-}v_c\right)}\left[\left(1\text{-}2v_c\right)\frac{t}{a}\ln\left(\frac{1+\left(d/a\right)^2}{\left(d/a\right)^2}\right) - \frac{\left(d/a\right)^2}{1+\left(d/a\right)^2}\right]$$

where $d$ is the film thickness and $a$ is the radius of the contact area estimated following the Oliver and Pharr's method as shown in **Figure 2**a. Analogously, for the effective Poisson's ratio $v_c$, we use the averaging formula[51]:

$$v_c = 1 - \left[\frac{\left(1-v_s\right)\left(1-v_f\right)}{1-\left(1-I_1\right)v_f - I_1 v_s}\right] \tag{4}$$

for which the weighting function $I_1$ reads:

$$I_1 = \frac{2}{\pi}\arctan\left(\frac{d}{a}\right) + \frac{1}{\pi}\frac{d}{a}\ln\left(\frac{1+\left(d/a\right)^2}{\left(d/a\right)^2}\right)$$

The value of $\nu_c$ is derived from Eq. (4) by assuming the Poisson's ratio of the azobenzene containing polymer film $\nu_f = 0.38$ for dark condition and $\nu_f = 0.45$ under illumination[30], while also assuming the Poisson's ratio for the glass of $\nu_s = 0.22$. The value of $G_c$ can be derived from the measured indentation modulus M as explained in section 3:

$$G_c = \frac{E_c}{2(1+\upsilon_c)} = -\frac{1}{2(1+\upsilon_c)} \frac{(-1+\upsilon_c^2) M \, E_i}{E_i - M + M\,\upsilon_i^2}$$

**Table 3** reports the estimated elastic modulus of the film $E_f$ for the azobenzene-containing polymer for the three states. It is interesting to notice that the effect of the substrate on the measured composite stiffness becomes more important under illumination as $I_0$ reduces from 40% before illumination down to 31 % under optical excitation. The reason is that, while changing the mechanical properties of the azo-polymer, light also affects the size of the film volume probed by nanoindentation. We conclude that the film modulus $E_f$ slightly reduces under illumination (from 8.7 GPa to 7.4 GPa) in spite of the increase in M due to the higher proximity effect of the glass substrate. Furthermore, it is important to mention that our study is limited to a fixed loading/unloading rate of 10 $\mu$N s$^{-1}$, while the results may depend on the strain rate sensitivity of the azobenzene containing polymer, as shown by Vapaavuori *et al.*[37] Thus, the estimated values of $E_f$ shall be taken as a first order approximation and further tests at different loading rates and advanced finite element analysis may be required to confirm these values.

| State | $G_s$ [GPa] | $G_f$ [GPa] | $G_c$ [GPa] | $E_f$ [GPa] | $v_s$ [-] | $v_f$ [-] | $v_c$ [-] | $d/a$ [-] | $I_0$ [%] | $I_1$ [%] |
|---|---|---|---|---|---|---|---|---|---|---|
| Initial | 34.7 | 3.2 | 6.9 | 8.7 | 0.22 | 0.38 | 0.34 | 0.90 | 40 | 70 |
| Excited | 34.9 | 2.6 | 7.0 | 7.4 | 0.22 | 0.45 | 0.38 | 0.70 | 31 | 64 |
| Relaxed | 34.7 | 3.2 | 7.0 | 8.9 | 0.22 | 0.38 | 0.34 | 0.92 | 41 | 70 |

**Table 3.** Values of the mechanical parameters of the 250 nm thick PMMA-DR1 film deposited on a glass substrate, deduced from the measurements presented in Figs. 5-7.

Analogously, one can estimate the hardness of the film $H_f$ from the measured hardness $H_c$ and the hardness of the substrate $H_s$ by applying the Bhattacharya and Nix's law[54] of mixture:

$$H_c = H_s\left(1 - I_H\right) + H_f I_H, \text{ where } I_H = \exp\left[-\frac{Y_f}{Y_s}\frac{E_s}{E_f}\left(\frac{h}{d}\right)^2\right].$$  (5)

The quantities $Y_f$ and $Y_s$ are the material yield stresses of the film and substrate, respectively. Since the only unknown is the ratio of $Y_f/Y_s$, we can plot the estimated $H_f$ according to Eq.5 as a function of the yield strength ratio $Y_f/Y_s$ (**Figure10**).

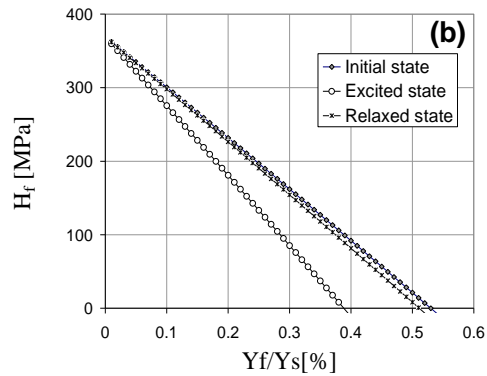

**Figure 10.** Estimated hardness $H_f$ of the film as a function of the yield strength values for the three illumination configurations, according to the mixture law proposed by Bhattacharya and Nix.[54]

Within the limit of validity of the equation (5), which is $Y_f/Y_s < 0.5\%$, the estimated hardness $H_f$ of the film under illumination is significantly smaller than the ones in the initial and relaxed cases. This result confirms our previous conclusion and rules out any substrate proximity effect in the reduction of H under illumination.

## 4.2 Creep recovery

To determine the reversibility of the creep deformation, one needs to compare the residual penetration depth $h_r$ reached after full unloading at the end of the *Experiment 2* procedure with the plastic penetration depth $h_p$ developed at the end of the loading phase. As discussed in Sec. 3, the latter can be obtained by assuming that the elastic modulus is constant during loading and unloading, so that the elastic penetration depth $h_e$ for a conical indenter can easily be deduced from Eq.2.[55] The creep recovery can be estimated by the parameter ω defined as $\omega = \dfrac{h_{max} - h_r}{h_{max} - h_p}$

| Variable | Initial | Excited |
|---|---|---|
| $h(\tau_L)$  [nm] | 62 | 68 |
| $h_e(\tau_L)$  [nm] | 23 | 24 |
| $h_p(\tau_L)$ [nm] | 39 | 44 |
| $h_{max}(\tau_{L+}\ \tau_H)$  [nm] | 74 | 85 |
| $h_r\ (\tau_{L+}\ \tau_{H+}\ \tau_{U)}$ [nm] | 53 | 69 |
| Creep recovery ω [%] | 60 | 39 |
| Dissipated energy γ [%] | 70% | 91% |

**Table 4**. Measured values of the different penetration depths for the three states of illumination.

When $h_r = h_p$, $\omega = 100\%$, which means that the elastic deformation developed during the holding phase is fully recovered after unloading. **Table 4** shows the values of $h_{max}$, $h_r$, $h_p$ and $h_e$ deduced from of the mean h-t curves for the three states with indication of the creep recovery parameter. The light illumination notably reduces the creep recovery from 60% to 39%, which corroborates the visco-plastic behavior of the polymer film under illumination, while it increases the dissipated energy from from 70% to 91%.

**4.3 Material viscosity changes**

To model the creep recovery, we apply the Burgers model as it can include both the reversible creep part (i.e., Kelvin-Voigt with $E_2$, $\eta_2$) and the irreversible part (i.e., Maxwell device with $E_1$, $\eta_1$) (Eq.1). **Figure 12** shows the comparison between the measured and the simulated creep curves of the experimental phase 2 of the creep recovery. The best fitting parameters of the Burgers model are listed in **Table 5.** In order to accurately reproduce the increase in the irreversible creep observed under light, the estimated irreversible viscosity $\eta_1$ diminishes from 1595 to 149 GPa.s as light is turned on, whereas no significant variation of the reversible viscosity $\eta_2$ is obtained. This diminution of the film viscosity under illumination is less important than the one previously reported using a significantly different experimental configuration.[30,33] Interestingly the effect of light is also to shorten the characteristic time of the irreversible viscosity $t_1$, which passes from 0.7 to 0.07 s. The effective modulus $E_{eff}$ of the film is simply estimated from the two springs in series.

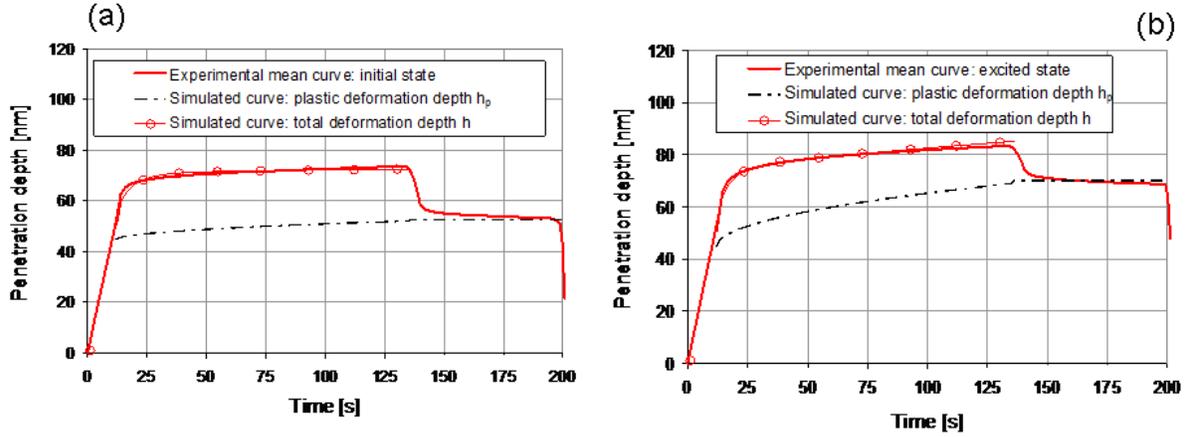

**Figure 11.** *Comparison of the simulated creep penetration depth with the measured curves in the initial state (a) and excited state (b) for the creep recovery test (Experiment 2). In the simulation the elastic part is described by the Burgers model and the irreversible plastic part by a Maxwell device (see Fig 3b).*

| State | $h_p(\tau_L) / h(\tau_L)$ [%] | $E_1$ [GPa] | $E_2$ [GPa] | $\eta_2$ [GPa.s] | $t_2 = \eta_2/E_2$ [s] | $\eta_1$ [GPa.s] | $t_1 = \eta_1/E_1$ [s] | $E_{eff}^{-1} = E_1^{-1} + E_2^{-1}$ [GPa] |
|-------|------|------|------|------|------|------|------|------|
| Initial | 88% | 2271 | 1.5 | 13.3 | 8.6 | 1595 | 0.70 | 1.5 |
| Excited | 88% | 2274 | 1.1 | 7.7 | 7.0 | 149 | 0.07 | 1.1 |

**Table 5.** Calculated parameters extracted from the Burgers viscous model (Eq. 1 et 3-a-b).

## Conclusion

In the present work we have investigated the variation of the mechanical properties of azobenzene containing polymers under light excitation by means of instrumented statistical nanoindentation measurements. With the state-of-the-art equipment used in this study, we have operated in a load range of a few tens of µN reaching corresponding penetration depths in the hundred of nm range, which is well-adapted to the study of thin polymer films. These experimental conditions are comparable with the conditions achieved with AFM-based nanoindentation, but with the advantage of independently control the load and the penetration depth. Moreover, quantitative measurements basically rely on the modeling of the nanoindenter tip, rather than on the several parameters needed to model the AFM cantilever deflection.

We performed three types of experiments: P-h loading experiments under illumination and in the dark, creep-recovery measurements and dynamical penetration measurements under light switching and constant load. Loading experiments show that optical excitation induces a significant increase in the hardness H (+38%) and in the creep C (+98%), while it has only a slight effect on the indentation modulus M (+8%). When the effect of the glass substrate is filtered out, the Young modulus of the film is found to significantly decrease by 15% under illumination. The optically induced modification of the material mechanical properties is mostly reversible: after light is turned off M, H and C retrieve their initial value.

The change in the Young modulus indicates that the light excited material exhibits a pronounced viscoplastic behavior. This trend is confirmed by the creep-recovery experiments, which show an increase in the residual deformation $h_r$ by 23% and an increase in the dissipated energy $\gamma_p$ by 30% when light is turned on. By applying simple Burgers viscous model, which leads to separate the reversible and the irreversible creep contributions, the irreversible part of the viscous deformation is found to increase by an order of magnitude under light excitation.

Finally, while applying a constant load, turning on the light induces a stiff increase in the penetration depth, which corresponds to the building up of the dynamical photostationary equilibrium between the trans and cis azobenzene molecule isomers. This is followed by a slow almost linear creep, which is likely related to the photoinduced mass motion usually observed in azo-materials.[56] When light is turned off, while creep is expected to stop and even to give rise to an elastic recovery, a surprising increase in the penetration depth is instead observed. It may be attributed to the reorganization of the polymer chains caused by the thermally activated back-transition of the cis molecules to the trans state. Further investigation of this phenomenon is in progress.